%% file: ms.tex
\documentclass[preprint]{vgtc}               

\ieeedoi{10.1109/VISUAL.2019.8933706}

\ifpdf
  \pdfoutput=1\relax                   
  \usepackage{graphicx}                
  \DeclareGraphicsExtensions{.pdf,.png,.jpg,.jpeg} 
\else
  \usepackage{graphicx}                
  \DeclareGraphicsExtensions{.eps}     
\fi%

\graphicspath{{figures/}{pictures/}{images/}{./}} 

\usepackage{microtype}                 
\PassOptionsToPackage{warn}{textcomp}  
\usepackage{textcomp}                  
\usepackage{mathptmx}                  
\usepackage{times}                     
\usepackage{cite}                      
\usepackage{tabu}                      
\usepackage{booktabs}                  

\input{files/preamble.tex}

\onlineid{1265}

\vgtccategory{Research}

\title{Slope-Dependent Rendering of Parallel Coordinates \\to Reduce Density Distortion and Ghost Clusters}

\author{David Pomerenke, Frederik L. Dennig, Daniel A. Keim, Johannes Fuchs, and Michael Blumenschein \\ \scriptsize University of Konstanz, Germany \\ \scriptsize  firstname.lastname@uni-konstanz.de}


\input{chapters/0-teaser.tex}

\abstract{
\input{chapters/0-abstract.tex}
} 

\begin{document}



\firstsection{Introduction}

\maketitle


\input{chapters/0-body.tex}

\acknowledgments{We thank the anonymous reviewers for their feedback. Funded by the Deutsche Forschungsgemeinschaft (DFG, German Research Foundation) within the projects A03 of TRR 161 (Project-ID 251654672) and Knowledge Generation in VA (Project-ID 350399414).}

\bibliographystyle{abbrv-doi}

\bibliography{ms}

\end{document}

%% file: files/preamble.tex
\usepackage[usenames,dvipsnames,table]{xcolor}
\usepackage{color}
\usepackage{graphicx}
\usepackage{capt-of}

\usepackage{amssymb}
\usepackage[nointegrals]{wasysym}

\usepackage{xspace}

\usepackage{enumitem}

\usepackage{hyperref}

\usepackage{tabu}

\usepackage{picinpar}

\usepackage[utf8]{inputenc}

\usepackage{microtype}
\usepackage{wrapfig}

\usepackage{subcaption}

\usepackage{hyphenat}

\usepackage[usenames,dvipsnames,table]{xcolor}
\usepackage{color}

\AtBeginDocument{

}

\usepackage{xspace}

\usepackage[leqno]{amsmath}
\usepackage{amssymb}
\newcommand{\leqnomode}{\tagsleft@true}

\newcommand{\subhead}[1]{\vspace{1pt} \noindent \textbf{#1}}

\usepackage{pgfplots}

\usepackage{lettrine}
\usepackage{wrapfig}

\usepackage{multirow,array,float}

\usepackage{soul}

%% file: chapters/0-teaser.tex
\teaser{
  \label{fig:top}
  \vspace{-0.5em}
  \centering
    
    \begin{tabular}{ cccc }
      \includegraphics[width=.24\linewidth,height=.14\linewidth]{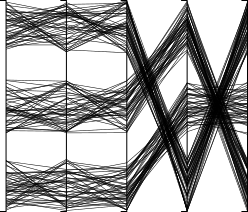} &
      \includegraphics[width=.24\linewidth,height=.14\linewidth]{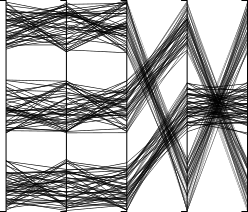} & \includegraphics[width=.24\linewidth,height=.14\linewidth]{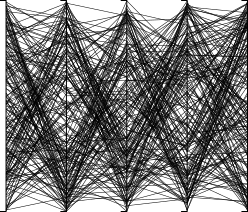} &
      \includegraphics[width=.24\linewidth,height=.14\linewidth]{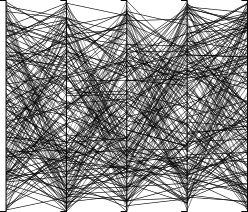} \\
      \small (a) Regular rendering & \small (b) Slope-dependent rendering & \small (c) Regular rendering & \small (d) Slope-dependent rendering
    \end{tabular}
    \vspace{-0.3em}
    \caption{\textbf{Comparison of regular parallel coordinates with our slope-dependent polyline rendering.}
    Parallel coordinates face two problems, which are inherent in the technique: 
    (a)~depicts three clusters of the same diameter and size across all dimensions. \emph{Diagonal changes of the clusters are visually more prominent}, as diagonal lines are rendered more closely. 
    (c)~shows 200 data points of uniform random clutter/noise in all dimensions.
    \emph{Zig-zag clusters are visible} as diagonal lines and are perceived as clusters, although there are no such clusters in the data (ghost clusters). 
    We propose to render each line segment based on its slope between two axes. 
    As a result, clusters are not distorted by their shape (b), and the ghost clusters effect is reduced (d).}
    \label{fig:teaser}
}

%% file: chapters/0-abstract.tex
Parallel coordinates are a popular technique to visualize multi-dimensional data. However, they face a significant problem influencing the perception and interpretation of patterns. The distance between two parallel lines differs based on their slope. 
Vertical lines are rendered longer and closer to each other than horizontal lines.
This problem is inherent in the technique and has two main consequences: (1) clusters which have a steep slope between two axes are visually more prominent than horizontal clusters. (2) Noise and clutter can be perceived as clusters, as a few parallel vertical lines visually emerge as a ghost cluster. Our paper makes two contributions:  First, we formalize the problem and show its impact. Second, we present a novel technique to reduce the effects by rendering the polylines of the parallel coordinates based on their slope: horizontal lines are rendered with the default width, lines with a steep slope with a thinner line. Our technique avoids density distortions of clusters, can be computed in linear time, and can be added on top of most parallel coordinate variations. To demonstrate the usefulness, we show examples and compare them to the classical rendering.

%% file: chapters/0-body.tex
\input{chapters/1-introduction.tex}
\input{chapters/2-related-work.tex}
\input{chapters/3-geometry.tex}
\input{chapters/4-solution.tex}
\input{chapters/5-discussion.tex}
\input{chapters/6-conclusion.tex}

%% file: chapters/1-introduction.tex
Parallel coordinates plots (PCPs)~\cite{DBLP:journals/vc/Inselberg85} are a well-researched visualization technique for multi-dimensional data. 
Studies have shown that they can be learned easily by non-visualization experts~\cite{DBLP:journals/iwc/SiirtolaR06, DBLP:conf/iv/SiirtolaLHR09} and used in practice in various domains like finance~\cite{Alsakran:2010}, traffic safety~\cite{Fua:1999}, and network analysis~\cite{Stockinger:2005}. 
Compared to related techniques such as scatter plot matrices and projections, PCPs have the advantage to identify, explore, and understand patterns across multiple dimensions. 
Cluster identification is, among others, one of the most common tasks for parallel coordinates~\cite{andrienko2001constructing}. 

Every data record is represented by a single polyline, spanning across the different axes/dimensions of the dataset. 
Polylines running close together are considered a cluster as they have similar values across the dimensions. 
In \autoref{fig:teaser}~(a), we can see three clusters spanning across the dataset. 
Between dimensions 1--3, the clusters are \emph{horizontal}, meaning that the data values are approximately the same within all dimensions. 
Across dimensions 3--5, the clusters are \emph{diagonal}, changing their values and cluster center, and have a steep \emph{slope}. 
We can easily see a general problem of the PCP technique: diagonal changes of clusters are visually more prominent than horizontal trends. 
Assuming all polylines have the same line thickness, there are two reasons for the problem: 
Diagonal lines need more area and pixels, and the actual space between parallel lines is smaller for diagonal clusters compared to horizontal ones. 
As a consequence, there is a density distortion of clusters based on the slope or angle of the cluster. 
A second problem, also based on these rendering artifacts, are so-called \emph{ghost clusters}. \autoref{fig:teaser}~(c) depicts a dataset with 200 points, randomly and uniformly distributed across all dimensions. 
One can ``see'' two zig-zag patterns indicating two clusters. 
However, the data does not contain any specific structure~-- in particular, no clusters. 
This problem is not only relevant in pure clutter (or noise) datasets but also influences the perception of clusters in datasets that contain a limited amount of clutter and noise along with relevant patterns. 
\emph{Ghost clusters} and distorted cluster density are related to human bias, but the core problem is based in the PCP technique. It can also occur in other variants of PCPs (e.g., different colors and transparency for lines, and edge-bundling).

We make two contributions: 
(1) we formalize the problem and show its impact.
(2) we propose a novel approach which renders each line segment based on the slope between two dimensions. 
Horizontal lines are rendered with the default line thickness. Diagonal lines are rendered thinner. 
Two examples are depicted in \autoref{fig:teaser}~(b) and (d).
The technique can be computed in linear time and applied on top of most PCP variations. The approach by Zhou et al.~\cite{zhou09} is closest to our work. It blends polylines based on their local neighborhood, which reduces the influence of noise but still suffers from the distortions caused by the over-emphasis on diagonal lines.

All material of this paper is available at \href{https://osf.io/sy3dv}{https://osf.io/sy3dv}.

%% file: chapters/2-related-work.tex
\begin{figure}[b]
    \centering
    \includegraphics[height=0.44\linewidth]{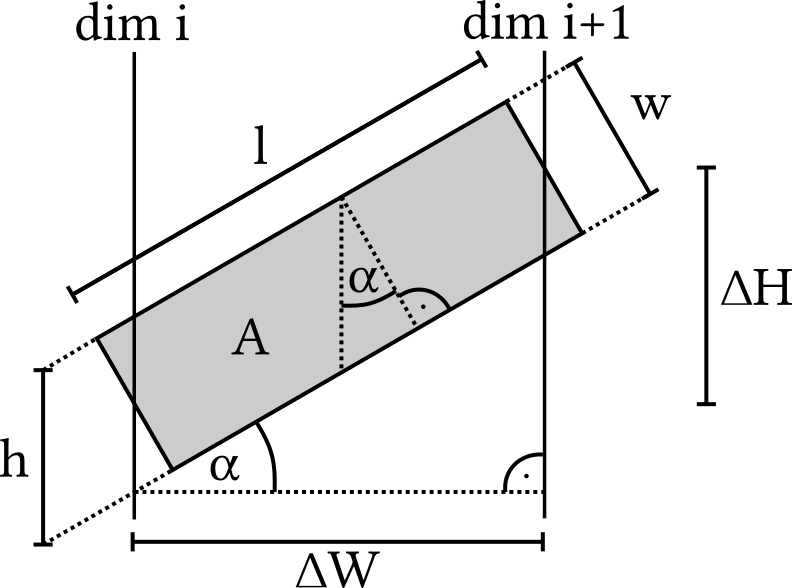}
    \caption{\textbf{Geometry of a polyline segment.} In regular
    PCPs, width $w$ is constant and length~$l$, height~$h$ and area~$A$ are dependent on $\alpha$.
    }
    \label{fig:sketch}
\end{figure}

\section{Related Work and Research Gap}\label{sec:related}
Plenty of approaches try to reduce clutter and highlight patterns in PCP generally. 
However, to the best of our knowledge, a formalization of the pattern distortion based on the polyline slope is missing, and none of the existing approaches specifically target this limitation.

\subsection{Sampling and Filtering Techniques}
The basic premise for the use of sampling (and filtering) techniques is that with less data, the degree of clutter and overplotting decreases, while the general structures, typically represented by many data records, remain in the PCP~\cite{DBLP:conf/eurographics/HeinrichW13}.
The taxonomy by Ellis \& Dix~\cite{ellis07} provides a categorization of clutter reduction methods, including sampling, filtering, and clustering, as well as visual techniques such as point size and opacity. 
Sampling often removes relevant data records or dimensions, and in this way reduces the truthfulness of the sampling concerning the dataset in its entirety. 
Our technique reduces clutter by counterbalancing the distortion artifact inherent to PCPs. It can be applied to a sampled or filtered subset of the data if the dataset exceeds the size visualizable in a PCP. Dependent on the data, our technique increases the amount of data displayable in a given PCP by deemphasizing diagonal polyline segments.

\subsection{Axes Reordering and Dimension Reduction}
Another approach to minimize clutter in PCPs is to reorder the dimension axes or reduce the number of displayed dimensions. For example, Pargnostics by Dasgupta and Kosara~\cite{dasgupta10} describes a set of quality metrics for PCPs which can be minimized or maximized (e.g., the number of line-crossings and parallelism). The authors also suggest the flipping of axes to reduce the number of line-crossings or diagonal clusters. 
The survey by Behrisch et al.~\cite{DBLP:journals/cgf/BehrischBKSEFSD18} discusses a large number of quality metrics as objective functions for axes reordering. 
Axes reordering, dimension reduction, and axes flipping can reduce ghost clusters by favoring horizontal structures. Depending on the data, however, it cannot be avoided entirely. Axes reordering is highly dependent on the data and analysis task. It is an orthogonal concept to our approach and can be combined with it.

\subsection{Density- and Cluster-based Rendering}
Clusters and other patterns can also be highlighted by density-distributed rendering. 
The general idea is to render PCPs as density distributions rather than individual polylines. Johansson et al.~\cite{johansson05} measure the density based on the number of overlapping polylines per pixel. This notion of density serves as input to a transfer function that allows highlighting areas according to their local density. Heinrich \& Weiskopf~\cite{heinrich09} apply the concept of continuous scatterplots~\cite{bachthaler08} to PCPs to derive a density model and thus interpolate the data. The resulting rendering is specifically useful for cluster identification. The work by Palmas et al.~\cite{palmas14} provides a different approach, which bundles edges according to class membership. The resulting bundles are rendered as polygonal strips.
Density- and cluster-based rendering may hide the underlying individual records and often require class labels to achieve a useful coloring or edge-bundling. 
While these approaches reduce clutter, they do not avoid the density distortion of clusters.

\subsection{Polyline Modifications}
A common technique is to modify the polylines of PCPs, specifically the overall line width, opacity, color, and shape. One example is the edge-bundling approach by Heinrich et al.~\cite{heinrich11}, which bundles polylines according to class membership and thus reshapes the line. The work by Zhou et al.~\cite{zhou09} called line splatting is most closely to ours. Line splatting is iteratively adjusting the opacity of lines based on the local neighborhood. Users can interactively change the degree of polyline and segment splatting. 
In contrast to Zhou et al.~\cite{zhou09}, our work tries to mitigate the visual distortions intrinsic to PCPs, such as the perceived density of clusters and the effect of ghost clusters. 

%% file: chapters/3-geometry.tex
\section{Problem Statement and Impact on PCP Patterns}

We formalize the line geometry of parallel coordinates and describe their effects on density distortions and ghost clusters. 

\subsection{Line Geometry in Parallel Coordinates}\label{sec:geometry}
In standard PCPs, a polyline segment has a constant line width ${w}\in\mathbb{R^+}$, also called thickness or stroke width. 
As depicted in \autoref{fig:sketch}, the slope of a segment is defined by the angle $\alpha\in [0, \frac{\pi}{2})$ 
between the horizon and the segment. $\Delta W$ denotes the space between the dimension axes and $\Delta H$  indicates the difference of data values.
In contrast to $w$, the line height is slope-dependent: $h = w \cdot \cos^{-1}(\alpha)$, with $\alpha = \tan^{-1}(\Delta H / \Delta W)$.
The area $A$ of a segment is defined as $h\cdot\Delta W$ and the length $l$ is defined as $\Delta W \cdot cos^{-1}(\alpha)$.

\begin{figure}[b]
    \centering
    \includegraphics[height=0.4\linewidth]{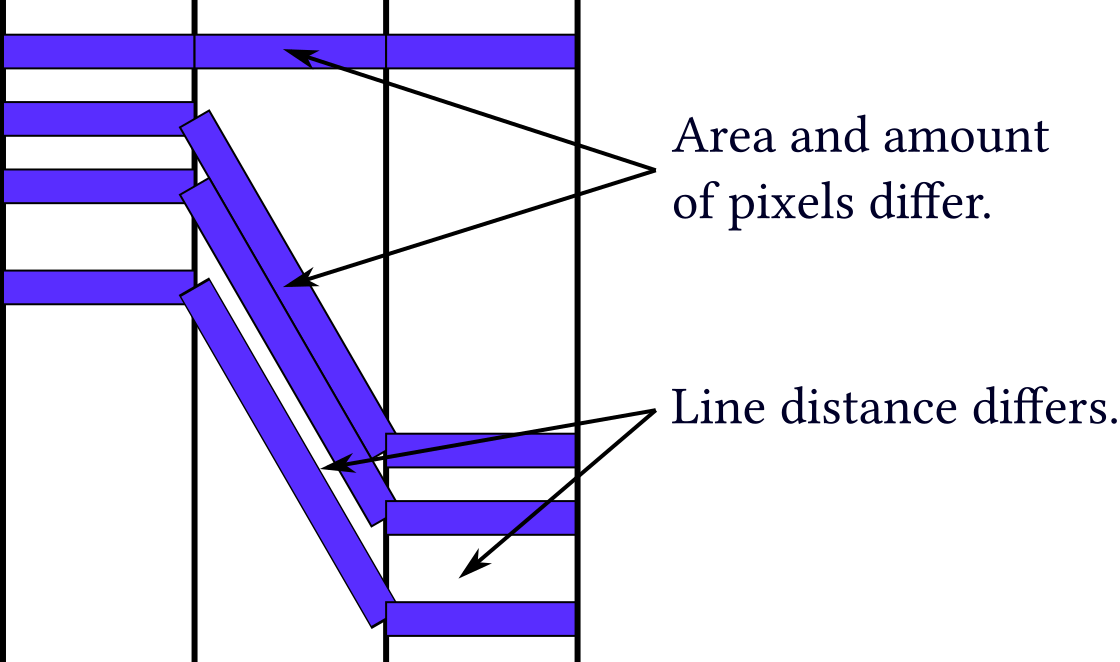}
    \caption{\textbf{Effect of angle $\alpha$ on PCP lines.} 
    (1) Diagonal lines have a \emph{higher line surface area} (=~more pixels) compared to horizontal lines. 
    (2) Diagonal lines have a \emph{smaller distance between lines}. 
    }
    \label{fig:line_effect}
\end{figure}

\begin{figure*}[t]
    \centering
    \begin{tabular}{ cccc } 
        \includegraphics[width=.22\linewidth]{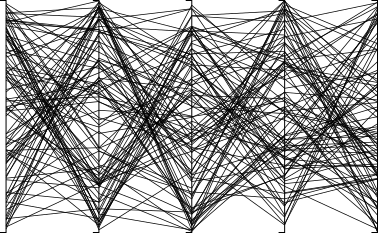} & 
        \includegraphics[width=.22\linewidth]{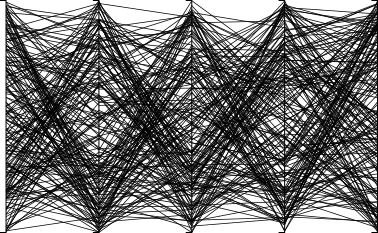} &
        \includegraphics[width=.22\linewidth]{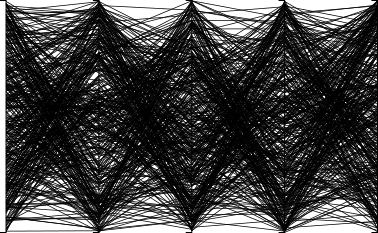} &
        \includegraphics[width=.22\linewidth]{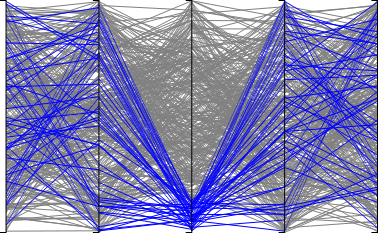} \\
        (a) $n=100$ & (b) $n=200$ & (c) $n=400$ & (d) $n=400$
    \end{tabular}
    \caption{\textbf{Ghost clusters in uniformly distributed random data points.} The number $n$ of polylines is increased from (a) to (c). (d) = (c) but the data points of a ghost cluster are highlighted to demonstrate that they are indeed uniformly distributed even though (c) indicates otherwise. }
    \label{fig:fake_patterns}
\end{figure*}

\subsection{Slope-dependent Distortion of Polylines}\label{sec:bias}
In parallel coordinates, \emph{horizontal clusters} correspond to a set of data points with a strong positive correlation in a subset of values across dimensions. 
Visually, these clusters have roughly horizontal cluster boundaries and only small line slopes. 
An example is depicted in dimensions 1--3 of \autoref{fig:teaser}~(a).
In contrast, \emph{diagonal clusters} correspond to data points with similar values within, but a strong variation between dimensions. 
Visually, these clusters have steep cluster boundaries and high line slopes. 
The last three dimensions in \autoref{fig:teaser}~(a) present examples.
Horizontal and diagonal clusters are not defined in a precise way, and there is a smooth transition between them. \emph{Visual cluster density} refers to the share of colored pixels within a line cluster. A large number of densely packed colored pixels induce a dense cluster and vice versa. 
The following effects characterize the emerging distortions influencing visual cluster density.

\begin{wrapfigure}[8]{r}{22mm} 
    \vspace{-10pt} 
    \hspace{-7mm}
    \includegraphics[height=27mm]{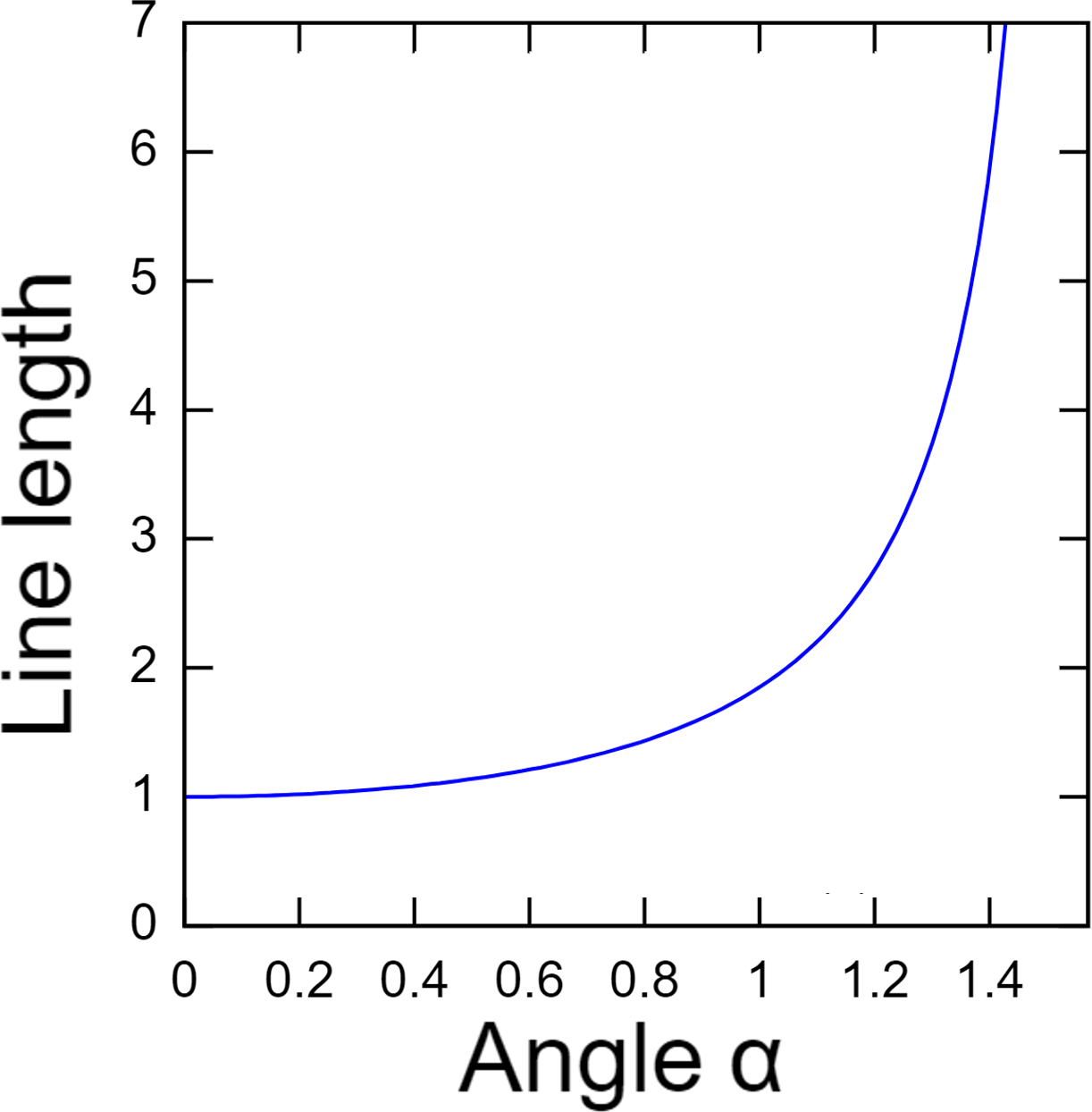}
\end{wrapfigure}

\subhead{Increase of Line Length and Area.} 
Line length $l$, line-height $h$ and line surface area $A$ depend on the angle $\alpha$, with the exponential relationship $A\propto h\propto l\propto cos^{-1}(\alpha)$ shown by the figure on the right. 
This dependency affects the perception of clusters. 
Large line slopes imply larger surface areas (=~more pixels, lower data-to-ink ratio~\cite{tufte86}) and therefore a more prominent line. The emphasis translates from lines to clusters, so that diagonal clusters are more noticeable than horizontal clusters. This effect is depicted in the top of \autoref{fig:line_effect}.

\subhead{Decrease of Line Distance.} 
Large line slopes in diagonal clusters reduce the space between lines and increases the perceived density of the cluster as lines may overlap, and the background vanishes. The orthogonal distance $d_\perp$ between two parallel lines is depends on the angle $\alpha$, with $d_\perp=d_h\cdot \cos(\alpha)$, where $d_h$ is the distance of the intersections of both lines with a dimension axis. This effect creates the perception that the lines are cohesive as shown in \autoref{fig:line_effect}.

\subsection{Visual Distortion of Cluster Densities}
The Gestalt law of proximity~\cite{koffka2013principles, ware2012information} indicates that the density of lines translates to a perception of cohesiveness and thereby enables users to recognize clusters in PCPs.
Classical PCPs put undue emphasis on diagonal clusters, which is facilitated by the \emph{increase of line lengths} and \emph{decrease of line distances}. 
This contradicts the data-ink ratio coined by Tufte~\cite{tufte86}, which describes the proportion of ink devoted to the actual data relative to the total amount of ink. Thus, it adds unnecessary distortion: Diagonal clusters are emphasized more than horizontal clusters. Classical PCPs, therefore, induce a systematically inaccurate perception of clusters, when the observer would expect that the visualization is inherently neutral in this respect. We can see the effect in \autoref{fig:teaser}~(a), where diagonal and horizontal clusters receive a significantly different emphasis.

\subsection{Ghost Clusters}\label{sec:clusters}
The rendering effects caused by the different slopes of the polyline segments can also produce artificial patterns in parallel coordinates plots. 
\autoref{fig:fake_patterns}~(a--c) show three PCPs with uniformly distributed random data points, i.e., there is no structure in the data. 
One can easily see that a \emph{zig-zag pattern}, alternating between high and low values is visually present. 
The corresponding polylines seem to be parallel and close together, forming two clusters. 
With an increasing number of data points, the ``clusters'' are perceptually stronger. 
In \autoref{fig:fake_patterns}~(d), we mark one apparent cluster and highlight its polylines across the different dimensions. 
One can see that the data is indeed randomly distributed and not forming a cluster across the dimensions. 
We define these visible, but non-existing patterns as \emph{ghost clusters}.
Ghost clusters are not only a problem of datasets with clutter or noise. Also, in structured datasets, ghost clusters can be present and influence the interpretation of the data.

%% file: chapters/4-solution.tex
\begin{figure*}[t]
    \centering
    \begin{tabular}{ cccc } 
        & Regular ($P=0$) & Adjusted ($P=1$) & Over-Adjusted ($P=2$) \\
        \rotatebox[origin=lB]{90}{\hspace{3mm}Synthetic Data} &
        \includegraphics[width=.30\linewidth,height=.15\linewidth]{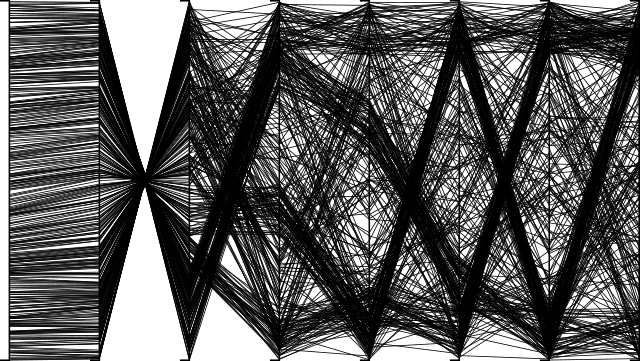} &
        \includegraphics[width=.30\linewidth,height=.15\linewidth]{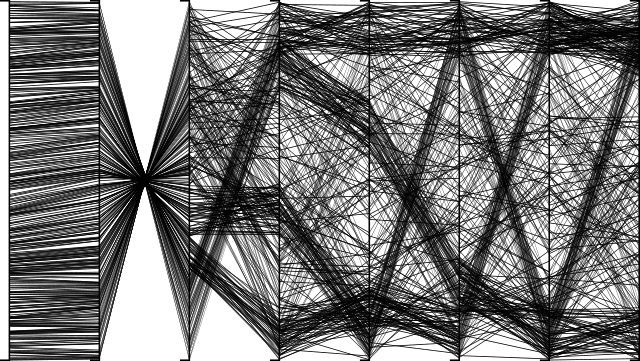} &
        \includegraphics[width=.30\linewidth,height=.15\linewidth]{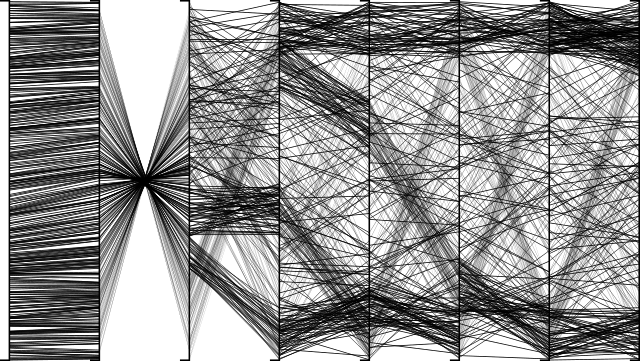} \\
        \rotatebox[origin=lB]{90}{\hspace{3mm}Uniform Noise} &
        \includegraphics[width=.30\linewidth,height=.15\linewidth]{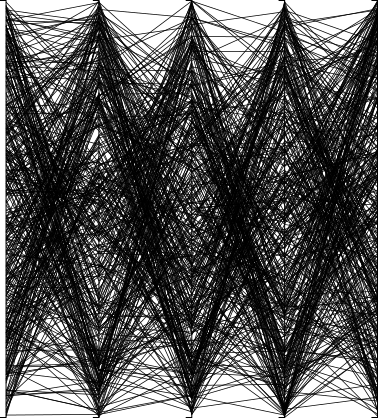} &
        \includegraphics[width=.30\linewidth,height=.15\linewidth]{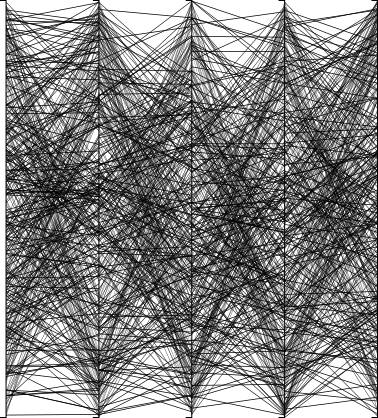} &
        \includegraphics[width=.30\linewidth,height=.15\linewidth]{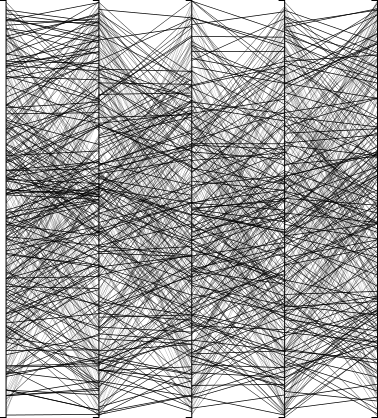}
    \end{tabular}
    \caption{
    \textbf{Effect of parameter $P$ on pattern visualization in synthetic data with uniformly distributed background noise, and in uniformly distributed random data only.} Regular rendering ($P=0$) significantly over-emphasizes diagonal clusters and causes the occurrence of ghost clusters. For $P=1$, all clusters are equally emphasized, and the effect of ghost clusters is strongly mitigated. For $P=2$ the distortion is reverted, and horizontal clusters are over-emphasized. Simultaneously, ghost clusters are further reduced.
    }
    \label{fig:bias}
\end{figure*}

\section{Slope-Dependent Rendering of Lines}\label{sec:solution}
To overcome the distortion of cluster densities and potential ghost clusters, we propose to render the polyline segments based on their angle $\alpha$. 
The general idea is to render horizontal lines with the default width and diagonal lines with a thinner line. 
As a result, we increase the space between vertical lines and decrease the surface area, i.e., the number of pixels to draw a line. 
In the ideal case, all line segments should end up with the same area and the same distance between the segments. 
To achieve the same area for all line segments, the width $w$ of the polyline segments needs to be scaled based on their length $l$. 
As the line length $l = \Delta W / cos(\alpha)$ is dependent on $\alpha$, the desired width $\omega$ also needs to depend on $\alpha$.
We interpret all lines as parallelograms with an equal and constant area $A$ and thus equal and constant side length $h\in\mathbb{R^+}$ which is \emph{independent} of $\alpha$ (\autoref{fig:sketch}). 
The height of this parallelogram corresponds to the desired $\alpha$-dependent width $\omega$, leading to $A = l \cdot \omega = \Delta W / cos(\alpha) \cdot (h \cdot cos(\alpha))$.
This results in the angle-dependent line width 
\begin{equation}
\omega=h\cdot \cos(\alpha)
\end{equation}

\noindent The angle-dependent width $\omega$ can be generalized, allowing us to weaken or strengthen the adjustment of the line width 
\begin{equation}
\omega =h\cdot \cos^P(\alpha)
\end{equation}
where parameter $P\in \mathbb{R}$ determines the adjustment strength. 
Our approach applies to pixel- and vector-based rendering techniques.


\subsection{Choosing the Adjustment Strength} 
$P=0$ corresponds to classical PCP rendering, where all lines have the same width. 
$P=1$ corresponds to rendering with equal line heights resulting in the same surface area $A$ for all polylines.
However, it does not fully correct the decreased line distances. 
Thus, we allow $P>1$ as over-adjustment to further compensate overplotting of lines with strong slopes. 
In particular, the parameter $P$ can be freely adapted to the degree of clutter, and the properties of the dataset. 
We want to highlight that our slope-dependent rendering can fully overcome the problem of different line surface area ($P=1$), but the issue of varying distance between polylines can only be reduced with $P>1$. 
Based on these geometric properties, we recommend $P=1$ for truthful representation. 
However, many properties of a PCP and dataset influence the quality of the rendering (see \autoref{sec:influence-parameters}), therefore an over-adjustment ($P>1$) may be necessary.
Our tests with various synthetic and real-world datasets showed that $P \approx 2$ is an upper bound for most applications.

In \autoref{fig:bias}, we apply our technique to a synthetic dataset and uniform random noise. We achieve a balanced emphasis of horizontal and diagonal clusters for $P=1$ and an over-emphasis of horizontal lines for $P=2$. Ghost clusters are also reduced for $P=1$ because their density is corrected. However, the effect of smaller line distance cannot be avoided, and ghost clusters are still visible. We can compensate for the line distance effect by over-adjusting the line area effect (e.g., $P=2$), nearly eliminating the ghost clusters, but introducing an over-emphasis of horizontal lines.

\subsection{Influence of PCP Properties and Parameters}\label{sec:influence-parameters}
The following parallel coordinates parameters influence the impact of ghost clusters and the distortion of cluster densities and should be taken into account when applying the slope-dependent rendering. 

\subhead{PCP Size, Axis Height and Spacing.} 
The overall size of a PCP has a direct impact on the axis height and spacing $\Delta W$ between the axes. 
Axis height and $\Delta W$ determine the range of $\alpha$: 
Long axes and tight spacing, caused by high-dimensionality, increase the angles and distort cluster densities and increase the likelihood of ghost clusters. 

\subhead{Default Line Width.} 
Manipulating the constant line-height $h$ influences the detail and the clarity of the PCP. Thick lines increase the problem of overplotting, in particular for diagonal lines and clusters. 
Thin lines are more distinguishable and therefore produce more salient visualizations. 
The result of the slope-dependent rendering depends on the default line width, typically determined by the user. 
The default width directly influences the area covered by each line segment. 
It is advisable to consider a manual adaptation of the constant line-height $h$ before applying a slope-dependent rendering.

\subhead{Data Volume.} 
The number of data records influences the visual representation a PCP and is strongly related to its size and the default line width. A high data volume visualized with a small PCP and/or a thick line width increases the problem of overplotting, but also the distortion of cluster densities and ghost clusters. 
For example, \autoref{fig:fake_patterns} shows how the dataset size increases the perception of ghost clusters. 
Therefore, these properties should be optimized for a given dataset before applying the slope-dependent rendering.

\begin{wrapfigure}[7]{r}{5mm} 
    \vspace{-12pt} 
    \hspace{-7mm}
    \includegraphics[width=10mm]{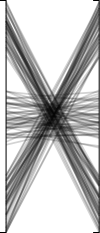}
\end{wrapfigure}
\subhead{Line Color and Transparency.} 
When no transparency is used, then the color of the polylines does not affect PCPs and therefore also not our approach. 
Transparency can be used to avoid clutter and overplotting but introduces another artifact, which negatively influences the perception of patterns. 
Crossing lines introduce a darker color, which may be interpreted as a cluster. 
Combined with the slope-dependent rendering, new ghost clusters may occur, while other patterns may vanish: 
Adjusting the transparency of lines based on their slopes, as opposed to the line width, is not useful.

%% file: chapters/5-discussion.tex
\section{Discussion and Future Work} \label{sec:discussion}

To test the effectiveness of our slope-dependent rendering, we implemented a tool which is available on our website\footnote{See \url{http://subspace.dbvis.de/pcp-adjustment} for the tool and \url{https://github.com/davidpomerenke/slope} for code and data.}.  
Users can upload their data, or try out various synthetic and real-world datasets, comparing the results of classical and slope-based rendering. 
During our testing with the implementation, we found out that our slope-dependent line adjustment technique performs well on various datasets, reduces ghost clusters, and counterbalances distortions. We also tested the impact of our approach with other patterns, such as positive and negative correlations (\autoref{fig:bias}). While positive correlations are not affected even with a large $P$ value ($P = 2$), the slope-dependent rendering influences the diagonal lines of negative correlation. We found that negative correlations also remain visible. However, the line representing data points at the ends of the dimension ranges are drawn with a small line width, making the visibility of this pattern susceptible to large $P$ values ($P = 2$).

Our approach can be combined with other techniques, such as axes reordering and dimension reduction, as they do not manipulate the polylines of a PCP. It can also be combined with polyline modifications like edge-bundling. However, the line width should then be calculated relative to the line length rather than the slope. As described above, various PCP properties generally influence the visual distortion and ghost clusters in PCPs. 
To achieve optimal results, these parameters should be optimized before the slope-dependent rendering is applied, and focus on the reduction of overplotting and the average angles of polylines. 

A careful selection of the parameter $P$ is necessary. 
The usefulness of a particular $P$ depends on many general PCP properties, as well as data characteristics such as the number of data records and dimensions. 
Therefore, $P$ cannot be determined fully automatically based on a fixed parameter. 
However, we envision an algorithm which measures the density distribution, overlapping, and distortion and automatically selects an appropriate $P$ to achieve a reliable representation of the data.
We want to address this algorithm as part of future work.
Furthermore, we want to evaluate the usefulness of our approach, in particular in comparison to other methods, by conducting a quantitative user study. 

%% file: chapters/6-conclusion.tex
\section{Conclusion} \label{sec:conclusion}
We formalize two general problems of parallel coordinates: 
The density of clusters are often distorted and non-existing ghost-clusters emerge. 
As a solution, we propose a novel rendering technique for the polyline segments: 
The line width is adjusted according to the angle of each line segment. 
Our method can be computed in linear time, depends on a single parameter, and can be combined with many existing parallel coordinates' variations. 